\documentclass[twocolumn]{aastex62} 

\usepackage{amsmath}
\usepackage{float}
\usepackage[colorinlistoftodos]{todonotes}

\newcommand{\sgra}{Sgr A$^*$}

\newcommand{\orcid}[1]{\href{#1}{\includegraphics[scale=0.035]{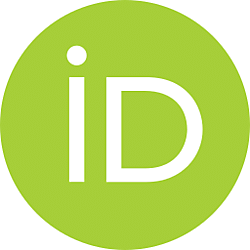}}}

\definecolor{boxback}{HTML}{DAE5F0}
\definecolor{boxframe}{HTML}{0F365B}
{
\end{tcolorbox}   
}

\newcommand{\eg}[1]{(e.g. \citealt{#1})}

\def\say#1{\textbf{[#1]}}

\received{December 19, 2019}
\revised{January 1, 2020}
\accepted{\today}
\submitjournal{ApJL}

\shorttitle{A nonthermal origin for \sgra's superflare}
\shortauthors{Guti\'errez et al.}

\begin{document}

\title{A nonthermal bomb explains the near-infrared superflare of Sgr A*}

\correspondingauthor{Eduardo M. Guti\'errez}
\email{emgutierrez@iar.unlp.edu.ar}

\author{Eduardo M. Guti\'errez \orcid{https://orcid.org/0000-0001-7941-801X}}
\affil{Instituto Argentino de Radioastronom\'ia (IAR, CCT La Plata, CONICET/CIC), C.C.5, (1984) Villa Elisa, Buenos Aires, Argentina}

\author{Rodrigo Nemmen \orcid{https://orcid.org/0000-0003-3956-0331}}
\affil{Universidade de S\~ao Paulo, Instituto de Astronomia, Geof\'{\i}sica e Ci\^encias Atmosf\'ericas, Departamento de Astronomia, S\~ao Paulo, SP 05508-090, Brazil}

\author{Fabio Cafardo \orcid{https://orcid.org/0000-0002-7910-2282}}
\affil{Universidade de S\~ao Paulo, Instituto de Astronomia, Geof\'{\i}sica e Ci\^encias Atmosf\'ericas, Departamento de Astronomia, S\~ao Paulo, SP 05508-090, Brazil}
\nocollaboration

\begin{abstract}
The Galactic center supermassive black hole, Sgr A*, has experienced a strong, unprecedented flare in May 2019 when its near-infrared luminosity reached much brighter levels than ever measured. We argue that an explosive event of particle acceleration to nonthermal energies in the innermost parts of the accretion flow---a nonthermal bomb---explains the near-IR light curve. We discuss potential mechanisms that could explain this event such as  magnetic reconnection and relativistic turbulence acceleration. Multiwavelength monitoring of such superflares in radio, infrared and X-rays should allow a concrete test of the nonthermal bomb model and put better constraints on the mechanism that triggered the bomb. 
\end{abstract}


\section{Introduction} \label{sec:intro}

At the center of the Milky Way lies Sagittarius A* (\sgra), a supermassive black hole (SMBH) with a mass of $M=4 \times 10^6 M_\odot$ located at a distance of 8.2 kpc \citep{Abuter2019}. Given its proximity, \sgra\ presents one of the best laboratories for studying the physics of black hole (BH) accretion flows \citep{Falcke2013}.
\sgra\ has been detected in most of the electromagnetic spectrum \eg{Dibi2014}. The extremely low accretion rate and low luminosity observed in its quiescent state ($L_{\rm bol}\sim 10^{36}~{\rm erg~ s^{-1}} \sim 2\times 10^{-9}L_{\rm Edd}$ where $L_{\rm Edd}$ is the Eddington luminosity) implies that the accretion flow is in a radiatively inefficient accretion flow (RIAF) state \eg{Yuan2014}. 

On top of the quiescent emission, \sgra\ also exhibits frequent flares in X-rays \eg{Neilsen2013, Ponti2015} and near-infrared (NIR) \eg{Genzel2003, Boyce2018}. About one X-ray flare is seen per day with a typical duration of a few tens of minutes \citep{Neilsen2013}. The brightest observed X-rays flares are $\sim 100$ times above the quiescent level \eg{Nowak2012}. The NIR flares are even more frequent.
X-ray flares usually follow the NIR ones after a few tens of minutes, but there are multiple NIR flares without a X-ray counterpart \eg{Eckart2006, Yusef-Zadeh2012, Ponti2017} (but see \citealt{Fazio2018}). Flares are also observed in mm and submm wavelengths \eg{Yusef-Zadeh2006, Stone2016}. They last from hours to days with amplitudes of $\sim 25\%$ the quiescent level \citep{Yusef-Zadeh2008, Fazio2018}. 

On May 2019, \cite{Do2019} observed an unprecedented NIR flare from \sgra---hereafter the ``superflare''---with the Keck telescope. The peak flux exceeded the maximum historical value by a factor of two and the light curve (LC) afterwards showed a factor of 75 drop in flux over a 2 hr time span. \cite{Do2019} suggested that an increase in the SMBH accretion rate $\dot{M}$ could be responsible for the superflare, possibly due to additional gas deposited by the passage of the G2 object in 2014 or a windy star such as S0-2 in 2018. Nevertheless, \cite{Ressler2018} argued that the effect of S0-2 on the RIAF structure should be negligible. This, combined with the fact that the S-star cluster has no known stars more massive than S0-2 close to Sgr A* spells trouble for the ``windy star'' scenario.

Here, we propose an entirely different scenario for the superflare which does not rely on an $\dot{M}$-increase: an explosive event of particle acceleration to nonthermal energies in the innermost parts of the accretion flow---a nonthermal bomb. This model explains quantitatively the NIR LC and makes testable predictions at other wavelengths.

\section{Model} \label{sec:model}

Our model for the emission involves a RIAF with populations of thermal and nonthermal electrons, following the height-integrated approach of \cite{yuan2003}. For simplicity, we assume that the dynamical structure of the flow (i.e. $\rho$, $\mathbf{v}$, $T$) does not vary with time, but we consider the possibility that an unspecified acceleration mechanism may change the number of particles following a nonthermal energy distribution. 

We take into account the presence of outflows by allowing the accretion rate to decrease with radius as $\dot{M}(r)=\dot{M}_{\rm max}(r/r_{\rm max})^s$ \citep{blandford1999}, with $s=0.25$.  We are only interested in the inner parts of the flow, so we only consider the accretion flow up to $r_{\rm max}=10^3r_{\rm S}$  where we set $\dot{M}_{\rm out} \approx 10^{-7} {\rm M}_\odot~{\rm yr}^{-1}$. The other parameters are the fraction of turbulent energy directly transferred to electrons $\delta=0.33$, the viscosity parameter $\alpha=0.1$, and the gas pressure to magnetic pressure ratio $\beta=9$.

\subsection{Quiescent state} \label{sec:steady}

To reproduce the quiescent state of the spectral energy distribution (SED), we assume that in each shell of the RIAF a fraction $\eta_{\rm q}=0.4 \%$ of the thermal energy density of electrons is in a nonthermal population with a broken power-law distribution:
\begin{equation}
N_{\rm q} (\gamma ;r) = \left \{
\begin{array}{ll}
K_{\rm q}(r)~ \gamma^{-p},      & \mathrm{if\ } \gamma_{\rm min} \le \gamma \le \gamma_{\rm c}, \\
K_{\rm q}(r)(p-1)\gamma_{\rm c}\gamma^{-(p+1)}, & \mathrm{if\ } \gamma_{\rm c} \le \gamma \le \gamma_{\rm max}.
\end{array}
\right.
\end{equation}
where $N_{\rm q}$ is the number density of electrons in the quiescent state, $\gamma$ is the electron Lorentz factor, $p$ is the spectral index at injection, $\gamma_{\rm c}$ is the ``cooling break'' Lorentz factor at which the the accretion time is equal to the cooling time, $t_{\rm acc}=t_{\rm cool}(\gamma_{\rm c})$ (cf. section \ref{sec:disc}) and $\gamma_{\rm min}$ and $\gamma_{\rm max}$ denote the minimum and maximum Lorentz factors respectively. 
We assume that thermal electrons radiate locally through synchrotron, bremsstrahlung and inverse Compton processes. 
For nonthermal electrons, we only consider synchrotron emission and adopt $p=3.6$. 

\begin{figure}
\centering
\includegraphics[width=1.0\linewidth]{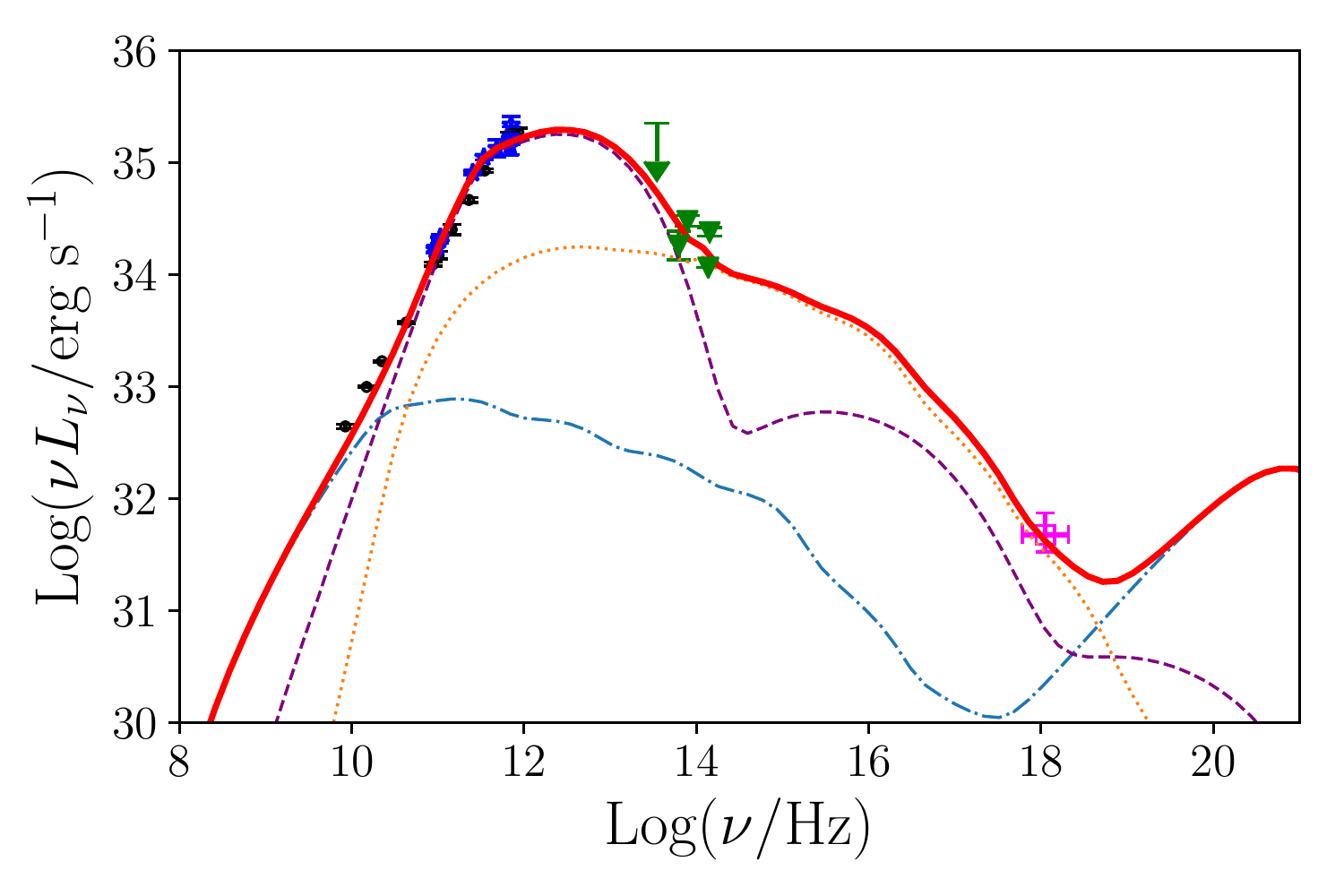}
\caption{Spectral Energy Distribution of Sgr A* in the quiescent state. 
The dotted line is the emission of nonthermal electrons from the inner parts of the flow ($r<15 r_{\rm S}$). The dashed line is the thermal synchrotron and inverse Compton emission. The dot-dashed line is the emission from the outer parts of the flow ($r>15 r_{\rm S}$), including thermal bremsstrahlung and nonthermal synchrotron. The solid line is the total emission.}
\label{fig:quiescent}
\end{figure}

Figure \ref{fig:quiescent} shows the quiescent state SED for the parameters given above. The observations are from \cite{liu2016} (radio, dark circles), \cite{shcherbakov2012} (radio, blue dots), \cite{schodel2011} (IR, green triangles), and \cite{roberts2017} (X-rays, magenta square). The submillimeter bump is due to thermal synchrotron, and the radio and IR excess are nonthermal synchrotron radiation.

\subsection{Flare}

Our model for flaring emission assumes that an unspecified process converts a fraction of electrons from the Maxwellian distribution to a nonthermal one during a short burst---a ``nonthermal bomb''. In Sec. \ref{sec:disc} we discuss about the possible physical mechanisms that might have produced such an event.

We consider that the burst occurs over an extended region ranging from radius $r_{\rm in}$ to $r_{\rm out}$. The injection function of nonthermal particles during a burst is 
\begin{equation}
\dot{N}_{\rm b}(\gamma,r;t) = \dot{N}_{\rm b}(\gamma,r)\delta(t),
\end{equation}
where $\dot{N}_{\rm b}(\gamma,r) = K_{\rm b}(r)\gamma^{-p_{\rm b}}$, and $K_{\rm b}(r)$ is determined imposing that at each shell a fraction $\eta_{\rm b} > \eta_{\rm q}$ of the thermal energy goes to nonthermal particles.
We follow the population while it is accreted onto the event horizon and compute the time evolution of the synchrotron emission. The transport equation that governs the evolution of this population is
\begin{widetext}
\begin{equation}    \label{eq:transport}
\frac{\partial N_{\rm b}(\gamma,r;t)}{\partial t} + \frac{1}{r^2}\frac{\partial}{\partial r} \Big [ r^2v(r)N_{\rm b}(\gamma,r;t) \Big ] + \frac{\partial}{\partial \gamma} \Bigg [ \Big (\frac{d\gamma}{dt}\Big )_{\rm syn}N_{\rm b}(\gamma,r;t) \Bigg ] = \dot{N}_{\rm b}(\gamma,r) \delta (t),
\end{equation}
\end{widetext}
where $d\gamma/dt(\gamma,r)$ is the rate of energy losses by synchrotron emission and $v(r)$ is the radial velocity of the flow. We solve equation \ref{eq:transport} by the method of characteristics.
There are five free parameters in the flare model: $\eta_{\rm b}$, the spectral index $p_{\rm b}$, $r_{\rm in}$, $r_{\rm out}$ and $t_0$ which is the time at which the burst occurs.

\section{Results} 

\begin{figure*}
\centering
\includegraphics[width=1.0\linewidth]{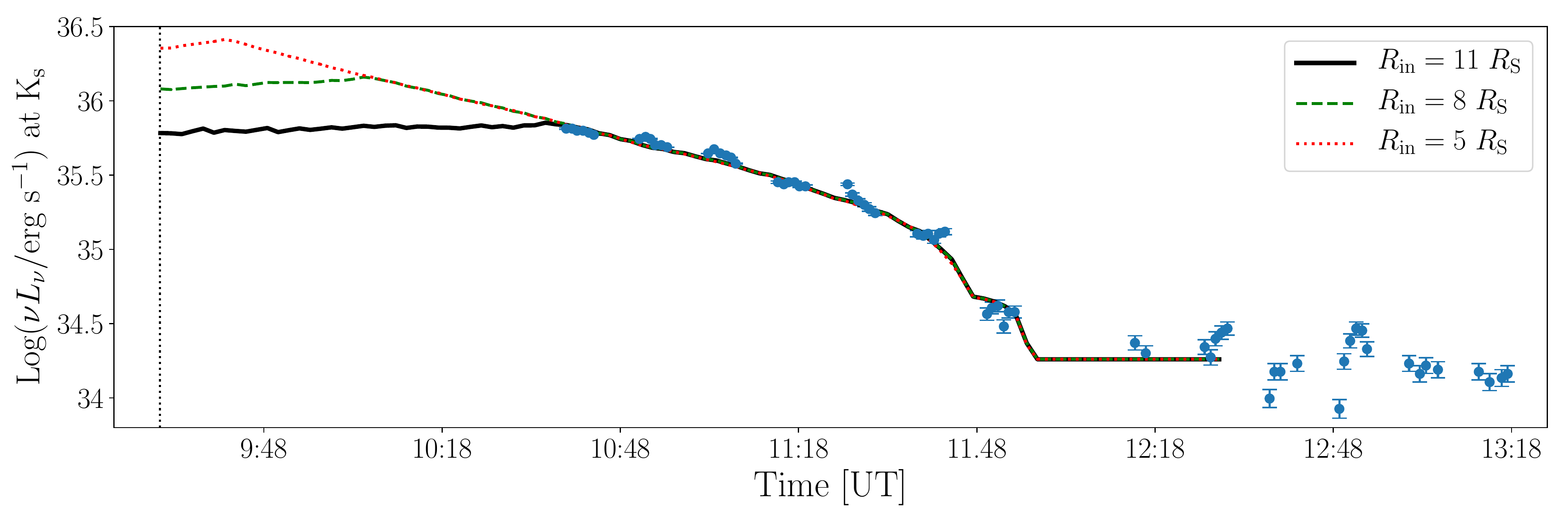}
\caption{Near-IR light curve of the superflare of Sgr A*. Points correspond to the Keck Telescope observations of \citet{Do2019} and lines indicate different nonthermal bomb models. The model parameters are $\eta_b = 0.25$, $p_{\rm b}=2.05$ and $r_{\rm out} = 16r_{\rm S}$, for three different values of $r_{\rm in}$. }
\label{fig:lc1}
\end{figure*}

Figure \ref{fig:lc1} contains the main result of this paper: we successfully explain the unprecedented bright state of Sgr A* observed in the NIR on May 2019 as an injection burst of nonthermal particles in the RIAF, which subsequently undergo radiative cooling as they get advected onto the black hole. The figure shows three models with different initial sizes of the burst region which reproduce well the decay in the NIR emission. The models reproduce the abrupt decrease in the flux in the last ten minutes of observations. This is interpreted as the accretion of the last nonthermal particles accelerated in the burst---those near $r_{\rm out}$ at $t=0$. 

\begin{figure}
\centering
\includegraphics[width=1.0\linewidth]{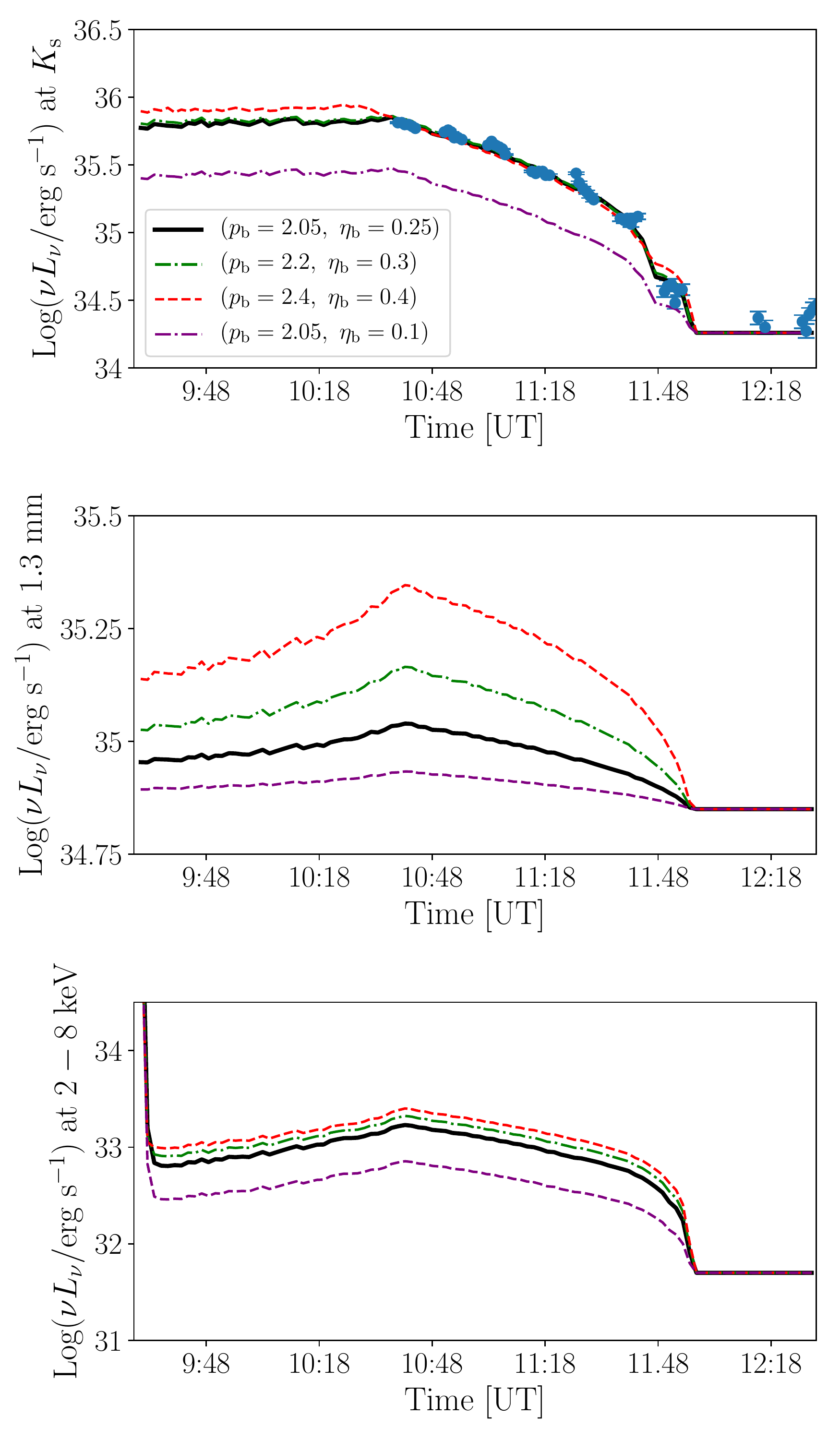}
\caption{Predicted flare emission at three wavelengths: near-IR (upper panel), $1.3$ mm (lower panel) and X-rays ($2-8$ keV; bottom panel). Three different values of the spectral index of the nonthermal distribution are displayed plus a model with the same spectral index as our fiducial model but with a lower value of $\eta_{\rm b}$. }
\label{fig:all_wavelengths}
\end{figure}

Our nonthermal bomb model predicts that the duration of the flare---determined by the accretion time---is the same across all wavelengths. The model also predicts that the slope of the LC following the initial burst depends on the wavelength. Both of these features are seen in Figure \ref{fig:all_wavelengths} which shows LCs in three different wavelenghts: NIR, $1.3$ mm (the Event Horizon Telescope wavelength) and 2-8 keV (the {\it Chandra} and \textit{XMM-Newton} energy band). The NIR LC is relatively insensitive to the slope of the electron energy distribution function, such that $L_{\rm NIR} \propto t^{-0.7}$. On the other hand, we find that the radio emission at mm-wavelengths depends modestly on the power-law index $p_{\rm b}$. This dependence can be approximated as $L_{\rm mm} \propto t^{0.4-0.25p_{\rm b}}$. The X-ray LC follows $L_{\rm X} \propto t^{0.4}$ and depends weakly on $p_{\rm b}$. Therefore, \textit{a campaign of multiwavelength monitoring of Sgr A*'s LC following a superflare in radio, NIR and X-rays should allow a concrete test of our model}.

Figure \ref{fig:all_wavelengths} also demonstrates that there are more than one combination of parameters capable of reproducing the NIR observations. For instance, the effect of the parameters $p_{\rm b}$ and $\eta_{\rm b}$ on the LC is degenerate: a change in any of these parameters affects only the total luminosity at the $K_{\rm s}$ band but does not modify the slope of the LC. This degeneracy can be broken by monitoring Sgr A* following the outset of the nonthermal bomb at other wavelengths. A change in $\eta_b$ only, leaving $p_{\rm b}$ fixed, modifies the total amount of energy in the bomb, and thus the luminosity at all times and wavelengths. This is shown in Figure \ref{fig:all_wavelengths}.

\section{Discussion} \label{sec:disc}

\subsection{Acceleration mechanism}

What is the mechanism responsible for the nonthermal bomb in Sgr A*? Black hole accretion flows are highly turbulent, highly magnetized, relativistic environments \eg{Porth2019}. Thus, plausible culprits are magnetic reconnection events and/or turbulence acceleration. In fact, magnetic reconnection has been invoked to explain the recurring IR and X-ray flares observed in Sgr A* \eg{Ball2018}. Shocks are unlikely because while being efficient at dissipating energy, they do not accelerate particles far beyond thermal energies \eg{Sironi2015}.

Numerical solutions of the Vlasov equation for astrophysical plasmas---i.e. particle-in-cell (PIC) simulations---are showing that: (i) Magnetic reconnection events with high magnetizations\footnote{The magnetization parameter is defined as $\sigma \equiv B^2/4\pi \rho c^2$, where $B$ is the magnetic field intensity and $\rho$ is the mass density---all quantities measured in the rest frame of the fluid.} of $\sigma \gtrsim 10$ lead to particles following power-law energy distributions with an index $p$ ranging from 1 to 2  \eg{Sironi2014, guo2014}, (ii) the presence of relativistic\footnote{Hereafter, by relativistic we mean that the mean magnetic energy per particle is larger than the rest-mass energy.} turbulence acceleration leads to a power-law index closer to $2$ \citep{comisso2019} and (iii) reconnection can deposit a large fraction (up to about $50\%$) of the dissipated energy in nonthermal electrons.

We have found that models with $p$ between $2$ and $2.5$ and $\eta_b \approx 0.25$ can account for the NIR flare evolution. Energy distributions with these parameters are consistent with having been produced within ten gravitational radii of the event horizon by either a magnetic reconnection event, or a reconnection event followed by relativistic turbulence acceleration.

For instance, according to the PIC simulations of \cite{Petropoulou2016} a lone reconnection event with $\sigma \approx 10$ should produce nonthermal electrons with the required values of $p$ and $\eta_{\rm b}$. Global GRMHD simulations such as those carried out by \cite{Ball2018} demonstrate that $\sigma$ is correlated with the plasma-$\beta$, $\beta \equiv P_{\rm gas}/P_{\rm magnetic}$. The values of $\sigma \gtrsim  10$ required to explain the superflare are only attained in configurations with high amounts of magnetic flux near the event horizon---i.e. the magnetically arrested disk (MAD) state---in regions of the accretion flow at which $\beta \sim 0.1$ \citep{Ball2018}. In our fiducial LC model, the total amount of magnetic energy involved in the burst is $\sim 3 \times 10^{40} \ {\rm erg\ s}^{-1}$. The MAD models of Ball et al. reach at most $\sim 10^{39} \ {\rm erg\ s}^{-1}$ for $\sigma \approx 10$, therefore a nonthermal bomb needs unusually large values of $B$---three times larger than the peak values of $B$ reached in MAD models. This would explain why superflares such as the one observed in May 2019 should be quite rare.

\subsection{Timescales} \label{subsec:timescales}

The relevant timescales for our problem are the electron cooling time and the accretion time. Interestingly, during the nonthermal bomb these timescales should be comparable. The synchrotron cooling time for an electron of Lorentz factor $\gamma$ is
\begin{equation}
	t_{\rm syn} \approx 7.74 \times 10^6 \Bigg ( \frac{B}{\rm 10~G} \Bigg )^{-2} \gamma^{-1}~{\rm s}.
\end{equation}
The cooling time corresponds to
\begin{equation}
	t_{\rm syn} \approx \Bigg ( \frac{B}{10~{\rm G}} \Bigg )^{-3/2}~{\rm h}.
\end{equation}
For magnetic fields of the order of $10~{\rm G}$, as appropriate for \sgra\ at  $\approx 10r_{\rm S}$, the cooling time is of the order of one hour. The accretion timescale is defined as $t_{\rm acc} = R/|v|$. Using the self-similar RIAF solution \citep{Narayan1994} we obtain a first-order estimate of this timescale as 
\begin{equation}
	t_{\rm acc} \approx 3 \alpha r^{3/2}~{\rm h}.
\end{equation}
For $\alpha=0.1$ and $r \approx 10$, $t_{\rm acc} \sim 10~{\rm h}$. 
In the models displayed in Figure \ref{fig:lc1}, the duration of the flare is determined mainly by the accretion time, but the slope also depends on the electron cooling. 
However, we find that a model only taking into account cooling 
with electrons remaining at a fixed distance from the hole---i.e. undergoing convective motion---also fits well the data. This shows that cooling can have an effect as important as accretion in our model.

\section{Summary}	\label{sec:summary}

Sgr A* has experienced a strong, unprecedented flare in May 2019 when its near-IR luminosity reached much brighter levels than ever measured. We have explained this superflare with a nonthermal bomb model, where an unspecified process accelerates over a very short time a small fraction of the electrons into a nonthermal distribution; these electrons subsequently cool and are advected onto the black hole. Besides explaining the NIR light curve, our model predicts that the radio and X-ray fluxes should decay over time in a similar fashion. In particular, the radio LC at mm-wavelengths is sensitive to the particle energy distribution and dissipation efficiency.

The nonthermal bomb detonated in a region spanning a length $5 R_S$ in the innermost parts of the accretion flow, and is likely due to a magnetic reconnection event involving unusually strong magnetic fields and high magnetization, i.e. $\sigma \gtrsim 10$, or such a reconnection event followed by turbulence acceleration. 

A multiwavelength monitoring of such superflares in radio, NIR and X-rays should allow a concrete test of the nonthermal bomb model and better constrain the mechanism that triggered the bomb. Future theoretical research should investigate the observational signatures of relativistic reconnection and relativistic turbulence acceleration using realistic magnetic field configurations appropriate for the SMBH in our Galactic Center, combining the tools of multidimensional GRMHD and PIC simulations.

\acknowledgments

We thank Reinaldo Santos de Lima for useful discussions and the anonymous referee for very constructive comments that led to improvements in the paper. E.G. thanks Gustavo Romero and Florencia Vieyro for useful discussions about relativistic processes in the vicinity of black holes.
This work was supported by the Argentine National Scientific and Technical Research Council (CONICET, grant PIP 2014-00338), the National Agency for Scientific and Technological Promotion (PICT 2017-0898), Funda\c{c}\~ao de Amparo \`a Pesquisa do Estado de S\~ao Paulo (FAPESP, grant 2017/01461-2) and Conselho Nacional de Desenvolvimento Cient\'ifico e Tecnol\'ogico (CNPq, grant 142320/2016-1).

\section*{ORCID iDs}

Eduardo M. Gutierrez \orcid{https://orcid.org/0000-0001-7941-801X} \url{https://orcid.org/0000-0001-7941-801X}

Rodrigo Nemmen \orcid{https://orcid.org/0000-0003-3956-0331} \url{https://orcid.org/0000-0003-3956-0331}

Fabio Cafardo \orcid{https://orcid.org/0000-0002-7910-2282} \url{https://orcid.org/0000-0002-7910-2282}

\bibliography{refs2,refs}

\end{document}